\documentclass [11pt]{article}

\def\draft{
\setlength{\oddsidemargin}{0pt}
\setlength{\evensidemargin}{0pt}
\setlength{\headsep}{0pt}
\setlength{\topmargin}{0pt}
\setlength{\textheight}{8.6in}
\setlength{\textwidth}{6.5in}
}

\draft

\usepackage {amssymb}
\usepackage{amsmath}
\usepackage {wasysym}
\usepackage {citesort}
\usepackage{color}
\usepackage[utf8]{inputenc}

\usepackage {graphicx}
\usepackage {subfigure}

\newcommand {\subplaatje} [2]
{
  \hfill
  \subfigure []
  {
    \includegraphics [#1] {#2}
    \label {#2}
  }
}

\newcommand {\eenplaatje} [3] []
{
  \begin {figure} [tb]
    \centering
    \includegraphics [#1] {#2}
    \caption {#3} \label {#2}
  \end {figure}
}

\newcommand {\plaatjesbreed} [3]
{
  \begin {figure*} [tb]
    \centering
    #1
    \hfill
    \mbox {}
    \caption {#2}
    \label {#3}
  \end {figure*}
}

\newcommand {\drieplaatjesbreed} [5] []
{
  \plaatjesbreed
  {
    \subplaatje {#1} {#2}
    \subplaatje {#1} {#3}
    \subplaatje {#1} {#4}
  } {#5} {#2+#3+#4}
}

\newcommand {\mathset} [1] {\ensuremath {\mathbb {#1}}}
\newcommand {\R} {\mathset {R}}

\newcommand {\eps} {\varepsilon}

\newcommand {\etal} {\emph{et al.}}
\newcommand {\seg} [1] {\ensuremath {\overline {#1}}}

\def\EXP{{\cal  E}}

\newtheorem{theorem}{Theorem}

\newtheorem{lemma}[theorem]{Lemma}

\newtheorem{obs}[theorem]{Observation}

\outer\def\proof{\vskip0pt plus.3\vsize\penalty-15
  \vskip0pt plus-.3\vsize\bigskip\vskip\parskip
  \noindent{\sc Proof.\enspace}}

\newcommand{\singlespace}{\def\baselinestretch{1.0}\Large\normalsize}
\newcommand{\doublespace}{\def\baselinestretch{1.5}\Large\normalsize}

\title {The dilation of the Delaunay triangulation\\
 is greater than $\pi/2$}

\author
{
  Prosenjit Bose\thanks
  {
    School of Computer Science, Carleton University. 5302 Herzberg Laboratories. 1125 Colonel By Drive, Ottawa, Ontario K1S 5B6, Canada {\tt jit@scs.carleton.ca}, research supported in part by NSERC and MRI.
  }
  \and Luc Devroye\thanks
  {
    School of Computer Science, McGill University, 3480 University Street, Montreal, Canada H3A 2K6. {\tt lucdevroye@gmail.com}
  }
  \and Maarten L\"offler\thanks
  {
    Department of Information and Computing Sciences, Utrecht University, the Netherlands, {\tt loffler@cs.uu.nl},   research  supported in part by the Netherlands Organisation for Scientific Research (NWO) through the project GOGO.
  }
  \and Jack Snoeyink\thanks
  {
    UNC Computer Science, Chapel Hill, NC, USA 27599--3175, {\tt \{snoeyink,verma\}@cs.unc.edu}
  }
  \and Vishal Verma\footnotemark[4]
}

\date {}

\index{Bose, Jit}
\index{Devroye, Luc}
\index{L{\"o}ffler, Maarten}
\index{Snoeyink, Jack}
\index{Verma, Vishal}

\begin {document}

  \maketitle

  \begin {abstract}
Consider the Delaunay triangulation $T$ of a set $P$ of points in the plane as a Euclidean graph, 
in which the weight of every edge is its length.  It has long been conjectured that the dilation in $T$ of any pair $p,p'\in P$, 
which is the ratio of the length of the shortest path from $p$ to $p'$ in $T$  over the Euclidean distance $\|pp'\|$, 
can be at most $\pi / 2\approx 1.5708$. In this paper, we show how to construct point sets in convex position with dilation $> 1.5810$ and in general position with dilation $> 1.5846$. 
Furthermore, we show that a sufficiently large set of points 
drawn independently from any distribution 
will in the limit approach the worst-case dilation for that distribution.
  \end {abstract}

\doublespace

  \section {Introduction}
In this paper we establish new lower bounds for the dilation of the Delaunay triangulation of points in the plane. 
We say that a graph $G=(P,E)$ on a set $P$ of points in the plane is a {\it Euclidean graph} 
if the weight of each edge $(p,q)\in E$ is the Euclidean distance $\|pq\|$.   
The {\it dilation} for any pair of points $p,p'\in P$ is the ratio of the length of a shortest path from $p$ to $p'$ in $G$ over the Euclidean distance $\|pp'\|$.  The \emph {dilation} of $G$ is the maximum dilation between any pair. 
The dilation of $G$ has also been called the {\it stretch factor} or {\it spanning ratio} of $T$, and the concept has been used to analyze routing algorithms for  networks~\cite{Bose2004273,gao05geometric,xlph04} and to define 
 \emph{spanners}~\cite{Annette2006,Farshi2005,langerman02computing,narasimhan00approximating,ns-gsn-07}: a $t$-spanner is a graph defined on a set of points such that the dilation between any two points is at most~$t$. 

A Delaunay triangulation of $P$ is a triangulation of the convex hull of $P$ in which the circumcircle of every triangle contains 
no  points of $P$ in its interior.  Note that each circumcircle will have at least the three vertices of the triangle 
on its boundary, and that if it has more points of $P$ on the boundary, the Delaunay triangulation is not unique.  
Our definition allows any triangulation of the points on a circle with empty interior to be called a Delaunay triangulation.

One of the first results in computational geometry on spanners was a proof of Chew's conjecture~\cite {c-pgagcg-89} that the Delaunay triangulation is a spanner in the plane.
Dobkin, Friedman and Supowit proved that the dilation of the Delaunay triangulation of any set of points in the plane is at most $(1+\sqrt 5)\pi/2<5.084$~\cite{dfs-dgaag-90}; the best upper bound known is $t=(4\sqrt3/9)\pi< 2.419$ by Keil and Gutwin~\cite{kg-cgwac-92}.  
At CCCG 2007 the first author posed, as an open problem~\cite {dor-op-08},  improving  dilation bounds for the special case of a set of points in convex position.  In this issue, Cui, Kanj and Xia establish an upper bound  of 2.33 
for points in convex position~\cite{ckx-oddt-09}, and we establish a new lower bound.

Until now, the only lower bound on the dilation for Delaunay triangulations was  Chew's  construction, which  achieves $\pi/2-\eps$, for any $\eps$,  by  sampling points $P$ uniformly on a unit circle~\cite {c-pgagcg-89}. Identify two antipodal points $p$ and $p'$, and triangulate $P$ by taking all edges nearly perpendicular to \seg {pp'}, as in Figure~\ref {circle}. Since all points are co-circular, any planar triangulation of $P$ is a valid Delaunay triangulation; alternatively, we could perturb the points to break the co-circularity, and make this the only triangulation.  The shortest path from $p$ to $p'$ via the triangulation follows the boundary of the circle, so its length approaches $\pi$. On the other hand, the Euclidean distance is clearly 2, so the dilation $t \to \pi / 2$.

    \eenplaatje {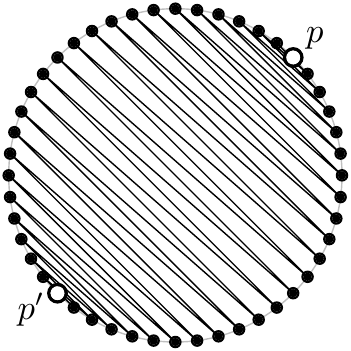} {The dilation of this Delaunay triangulation of a set of points on a circle approaches $\pi / 2$.}
    In their book on spanners, Narasimhan and Smid mention that ``it is widely believed that, for every set of points in $R^2$, the Delaunay triangulation is a $(\pi/2)$-spanner''~\cite {ns-gsn-07}.
We show here how to construct point sets in convex position whose Delaunay triangulation is not a  $(\pi/2)$-spanner, but in fact has dilation $>1.5810$.
We then modify the construction to create point sets not in convex position  whose Delaunay triangulation 
has slightly larger dilation; we do not yet know whether the maximum dilation for Delaunay triangulations can be 
approached by points in convex position. 
 Finally, based on these constructions, we prove that as you draw sufficiently many random points from any  single distribution, the dilation of the Delaunay triangulation approaches the worst-case dilation for that distribution.

  \section {Constructions with dilation greater than $\pi/2$}

 \eenplaatje {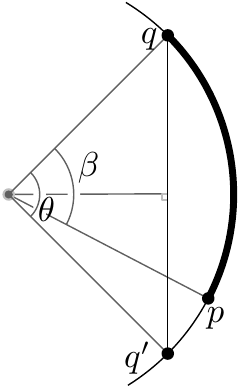} {Observation~\ref{obs:angle} compares two ways to go from $p$ to $q$: direct by arc of length $\beta$ or via $q'$ by arc+segment  of length $\theta-\beta+2\sin(\theta/2)$.  These balance if $\beta=\theta/2+sin(\theta/2)$.  When $qq'$ is a diameter, $\beta = 1+\pi/2$ radians.}

To construct  examples of points with dilation greater than $\pi/2$,
we use a simple observation about paths around sectors of unit circles that is illustrated in Figure~\ref{obs1}:
\begin{obs}\label{obs:angle}
Consider the segment of a unit circle defined by points $q$, $q'$ that subtend angle $\theta$, and place $p$ between them so that arc $pq$ has angle $\beta$.  Let $A$ be the path consisting of the arc from $p$ to $q'$ followed by the segment $qq'$. The arc from $p$ to $q$ around the boundary of this sector is shorter than the length of $A$ when $\beta<\theta/2+\sin(\theta/2)$.  
\end{obs}
\begin{proof}
The arc $pq$ has length $\beta$, while the arc $pq'$ and the segment $qq'$ together have length $\theta-\beta+2\sin(\theta/2)$. 
\end{proof}

    We are now ready to describe the construction for points in convex position.
    Form a convex region bounded by two unit semicircles having centers on the $x$ axis separated by distance $d$. As in Figure~\ref{ex-2base}, introduce points on these semicircles uniformly and identify two points $p$ and $p'$ at an angle of $\alpha$ from the $x$-axis.
    Next,  triangulate the semicircle with $p$ by adding chords to the convex hull in a way that ensures that any shortest path from $p$ to the endpoints of the semicircle follows the boundary of the circle.  One possibility is shown in Figure~\ref {ex-triangulation}.
Observation~\ref{obs:angle} ensures that the arc is the shortest path from $p$ to either endpoint of each  chord used to triangulate. 

   \drieplaatjesbreed  {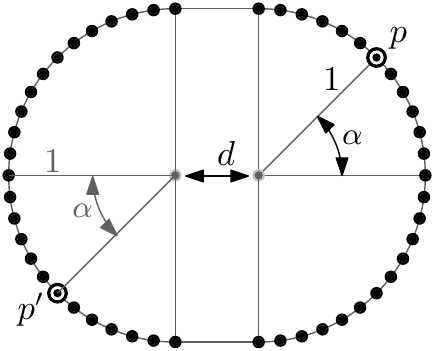} {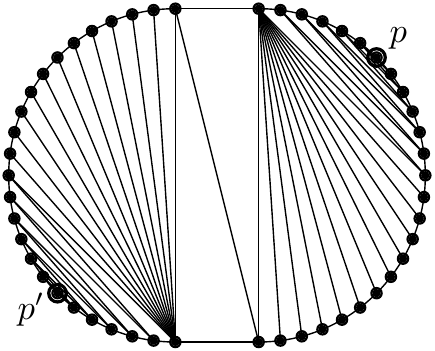}{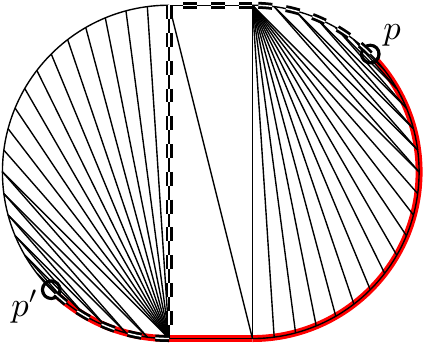} {(a) The basic construction consists of evenly sampled points on two unit semicircles with centers on the $x$ axis separated  by $d$. We mark two points $p$ and $p'$ that make an angle of $\alpha$ with the $x$-axis. (b) We choose the Delaunay triangulation in order to maximize the shortest path in the triangulation from $p$ to $p'$. (c) One locally shortest path (solid red) simply follows the boundary and has length approaching $\pi + d$; another (dashed) crosses the construction once and has length approaching $\pi + 2 - 2\alpha + d$.}

 In our triangulation there are two types of locally optimal paths from $p$ to $p'$, drawn solid and dashed in Figure~\ref {ex-2paths}. The first type, which follows the perimeter of the region (clockwise or counter-clockwise), has length $\pi + d$, since we walk around one semicircle and bridge the gap of width $d$. The other type, which crosses over via one of the vertical edges,  has length $2 \cdot (\pi/2 - \alpha)$ for the two circular arcs plus $2 + d$ for the straight parts, so $\pi + 2 - 2\alpha + d$ in total. Observation~\ref{obs:angle} ensures that any other path will be longer.
If we set $\alpha = 1$ radian, these two path types have equal lengths. 

Finally, we have to compute the Euclidean distance between $p$ and $p'$, 
which is $\ell=\sqrt {4 + d^2 + 4d\cos 1}$. 
The dilation approaches $t =  (\pi + d)/ {\ell}$.
As a function of $d$, $t$ remains above $1.5810528$ for $d \in [0.293, 0.294]$.
For $d=0.29$, we have $t > 1.581 > \frac \pi 2$.

\begin {theorem} \label {thm:1.581}
  There exists a set $P$ of points in convex position in the plane, such that the Delaunay triangulation 
of $P$ has a dilation of at least $1.5810$.
\end {theorem}
In this construction 18 points suffice to give a dilation $>\pi/2$ and 222 points give a dilation $>1.5810$.

    \eenplaatje {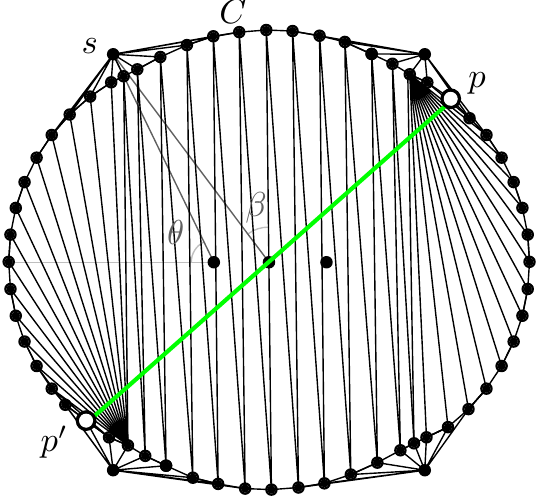} {Increasing the dilation by placing points on three circles, 
with four shield points. Two unit radius circles are separated by $d=0.58$, and the third circle, $C$,
of radius 1.15057 is centered at their midpoint.  Points $p$ and $p'$, placed by 
Observation~\ref{obs:angle}, have distance 2.4, and achieve a dilation of $1.5846$.}

If we allow points that are not in convex position, we can modify this construction, 
as illustrated in Figure~\ref{nonc-triangulation}, and increase the dilation slightly.  
The idea is to replace the two straight segments with points that lie on a common circle, $C$.  
The new triangulation edges inside the polygon will not be used by any shortest path 
from $p$ to $p'$; we need to prevent short-cuts outside the polygon by the edges 
needed to complete the triangulation there. 
At each of the four locations where a unit circle meets $C$, we add
a {\it shield} point $s$ on the line from the center of the unit
circle through the intersection, and draw a ray from $s$ through the center of $C$.  
We place
points densely on the arcs of all three circles, leaving  gaps in the angles
formed by the rays from each shield point~$s$. The position of $s$ on its line is chosen
so that the tangents from $s$ to the two circles form a path that is
just longer than the path that follows the circle boundaries.  The
triangulation outside the circles is completed by fans from each
shield point~$s$. 

For the best ratio, we separate the unit circles by $d=0.58$, and place circle $C$ of radius  $r=1.1507$ midway between their centers.  As depicted in Figure~\ref{nonc-triangulation}, the arcs of the unit circles are no longer semicircles, but subtend  $2\theta =2.2895$ radians; the two arcs of the larger circle $C$ each subtend $2\beta=\ 1.30432$ radians, with a shielded gap of $g =0.0065$.  When we place  $p$~and~$p'$ by Observation~\ref{obs:angle} we obtain a straight line of length $\ell=2.4$ and a dilation $(\theta+r\beta+g)/\ell >1.5846$.  
\begin {theorem} \label {thm:1.5846}
  There exists a set $P$ of points in the plane such that the Delaunay triangulation of $P$ has a dilation of $1.5846$.
\end {theorem}

\section {The dilation of the Delaunay for random point sets}
\def\RR{{\mathop{\rm I{\kern-1.60pt}R}}}
\def\IND#1{{1}_{{\left[ #1 \right]}}}
Because a small number of points may determine the dilation of a large Delaunay triangulation, we observe that dilation factors greater than $\pi/2$ are actually the rule, and not the exception, for sufficiently large sets of random points.  In this section we formally prove this for quite weak restrictions on the probability density function. 

Although the constructions of the previous section create degenerate point sets,  it is not hard to alter them slightly, for any finite number of points $k$, so that the points are in general position and the dilation is nearly the same.  Once the points are in general position, there is a value $\delta>0$ such that perturbing every point by at most $\delta$ leaves the combinatorial structure of the Delaunay triangulation unchanged.   (Abellenas \etal~\cite{ahr-stdt-99} provide an algorithm to compute $\delta$.)  The idea of this section is to show that whenever there is some finite probability that $k$ points chosen at random will create a Delaunay configuration with dilation greater than $\pi/2$, then when we choose $n\gg k$ points, we will, with high probability, find such a $k$-configuration.

The material below applies to general settings in which we seek a lower bound $L^*$,  that applies with high probability as $n\to \infty$, on a function $L_n$ on $n$ points in $R^d$ that is a maximum over configurations in products of intervals that are subsets of $R^d$. 
Assume the following two conditions:
\begin {enumerate}\itemsep=0pt
\item [(i)]
For each $n$, $L_n$ is scale and translation invariant, that is, 
$L_n (ax_1 + b, \ldots , ax_n + b) =  L_n (x_1,\ldots,x_n)$ for
all $a \not=0, b \in \R^d$.
\item [(ii)] 
There exists a value $L^*$ 
such that for all $\epsilon > 0$,
there exists a $\delta > 0$, integer $K \ge 1$, and points $x_1,\ldots,x_K \in [0,1]^d$,
such that if $B(x,\delta)$ denotes the ball of radius $\delta$ in  $\R^d$,
we have
$$
\inf_{z_1 \in B(x_1, \delta), \ldots, z_K \in B(x_K, \delta)} \inf_m \inf_{y_1, \ldots, y_m \not\in [0,1]^d} L_{K+m} (z_1,\ldots,z_{K}, y_1, \ldots, y_m) \ge L^* - \epsilon. \eqno{(1)}
$$
 
\end {enumerate}
By way of remark, the easiest way to satisfy (ii) is for $L_n$ to be a continuous function of its inputs;  condition (1) would then follow if the value $L^*$
were achieved for a certain point set. Unfortunately, the dilation of the Delaunay triangulation is not continuous; moving points by a small amount may induce a diagonal flip that changes the shortest path length.  Thanks to Abellenas, Hurtado, and Ramos~\cite {ahr-stdt-99}, however, we do have (ii) for non-degenerate point sets.
Briefly, (ii) is satisfied if there exists a small nonempty ball in the
neighborhood of (but not necessarily covering) a point at which $L^*$ is reached
on which $L_n$ is bounded from below by $L^* - \epsilon$.

To conclude this paper,   we prove a
general theorem on functions $L_n$ that satisfy our two conditions.
It essentially says that a copy of some pessimal construction can be
expected in a random point set under weak restrictions. The
probability density $f$ mentioned in the theorem is arbitrary 
(i.e., is the derivative of an arbitrary continuous cumulative distribution function).  
It can have unbounded
support and infinite peaks, and could possibly fail to be continuous at almost all~$x$.

\def\PROB{{\cal P}}
\def\isdef{\buildrel {\rm def} \over =}
\begin{theorem}\label{thm:rnd}
Let
$L_n\colon (\RR^d)^n\to \RR$  satisfy (i) and (ii) for a certain value $L^*$.
Let $X_1, \ldots , X_n$ be i.i.d.\  random vectors drawn from
a common density $f$ in $\RR^d$, then for every $\epsilon > 0$,
$$
\lim_{n \to \infty}
\PROB \left\{ L_n (X_1, \ldots, X_n ) \ge L^* - \epsilon \right\} = 1.
$$
\end{theorem}

\subsection{Preliminaries}

Let $S \subseteq [0,1]^d$ be a fixed set of positive volume.
Let $f$ be a density of $\RR^d$.
We say that $x$ is a Lebesgue point for $S$ if the density does not vary wildly in the neighborhood of $x$, expressed as:
$$
\lim_{r \downarrow 0} { \int_{x+rS} f(y) \, dy \over \int_{x+rS} dy } 
= \lim_{r \downarrow 0} { \int_{x+rS} f(y) \, dy \over r^d \int_{S} dy } 
= f(x).
$$
Thus, for a Lebesgue point $x$, the average density
value over a small ball centered at $x$ tends to the value of the density at $x$
as the radius decreases. A fundamental property of densities is that almost
all points are Lebesgue points.  
\begin{lemma}\label{lem:1}
For any density $f$, and any set $S \subseteq [0,1]^d$ of
positive volume, almost all $x$ are Lebesgue points for $S$.
\end{lemma}
\proof
See, e.g.,  Wheeden and Zygmund~\cite{wz-mi-77}, p.~108,
or de Guzman~\cite{dg-dir-75}. $\square$
\medskip

Unfortunately, we cannot conclude that the density is near uniform around Lebesgue points; 
convergence of averages does not imply uniform convergence. In particular,
it could be that at all $x$, the infimum of the density
over any such ball of any positive radius is zero, and the supremum is infinite.  
Therefore, additional technical work is needed.
We  extend the notion to sets
of points $x$ that are Lebesgue points for a finite collection
$S_1, \ldots , S_K$ from $[0,1]^d$. Clearly,
almost all $x$ are still in this set.
The next Lemma says that we can always find many
Lebesgue points that are far apart and have certain restrictions
on the densities.

\begin{lemma}\label{lem:2}
For any density $f$, and any sets $S_1,\ldots,S_K \subseteq [0,1]^d$ of
positive volume, there exist constants $0 < a < b < \infty$
such that for any integer $N > 0$,
we can find 
$\delta_0 > 0$,
$x_1,\ldots,x_N \in \RR^d$,
and constant $c > 0$
such that 
$b \ge f(x_i) \ge a$, 
the $L_\infty$ distance between each $x_i$ and $x_k$
is more than $c$, and 
	$$
	{1 \over \delta^d \lambda (S_j)}
	\int_{x_i + \delta S_j} f \in [ f(x_i)/2 , 2f(x_i) ],
        $$
for all $1 \le i \le N$, $0 \le j \le K$, $0 \le \delta \le \delta_0$,
	where $S_0 = [0,1]^d$.
\end{lemma}
\proof
The existence of the claimed point set is shown by the first moment
method.
Consider i.i.d.\ random points $X'_1, X'_2, \ldots$ drawn from $f$.
Choose $a$ and $b$ such that $\PROB \{ f(X'_1) \not\in [a,b] \} < 1/2$.
Let $X_1, \ldots, X_N$ be the first $N$ of the $X'_j$'s that have
$f(X_i) \in [a,b]$.  Let $T$ be the number of random vectors needed
to achieve this. Note that $\EXP \{ T \} \le 2N$,
as $T$ is stochastically smaller than a sum of $n$ geometric random variables
with parameter $1/2$.
With probability one, all $X'_j$'s, and thus
all $X_j$'s, are Lebesgue points for $S_0=[0,1]^d, S_1, \ldots, S_K$.
Thus, there exists $\delta_0$ (depending upon $N$, $f$, $X_1,\ldots,X_N$)
such that for $\delta \le \delta_0$,
for all $1 \le i \le N$, $0 \le j \le K$,
	$$
	{1 \over \delta^d \lambda (S_j)}
	\int_{X_i + \delta S_j} f \in [ f(X_i)/2 , 2f(X_i) ] ,
        $$
where $\lambda$ denotes Lebesgue measure.
The probability that the $L_\infty$ distance between $X_i$
and $X_j$ is less than $c$ for some $i \not= j$ is not larger than
$\EXP \{ T^2  M(c) \}$, by the union bound, where
$$
M(t) \isdef \sup_x \int_{x + t [0,1]^d} f(y) \, dy.
$$
As $\EXP \{ T^2 \} = O(N^2)$,
this is less than $1/2$
by choice of $c > 0$ because $M(c) \downarrow 0$ as $c \downarrow 0$.
We take such a $c$ (depending upon $N$).
Thus, by the first moment method,
there exists a $\delta_0, x_1,\ldots,x_N, c > 0$
such that
jointly $b \ge f(x_i) \ge a$, the $L_\infty$ distance between each $x_i$ and $x_j$
is more than $c$, and 
	$$
	{1 \over \delta^d \lambda (S_j)} \int_{x_i + \delta S_j} f \in [ f(x_i)/2 , 2f(x_i) ] ,
        $$
for all $1 \le i \le N$, $0 \le j \le K$, $0 \le \delta \le \delta_0$. $\square$
\medskip

\subsection{Proof of Theorem~\protect\ref{thm:rnd}}

We are now ready for the proof of Theorem~\ref{thm:rnd},  which 
proceeds by the second moment method.
Let $\epsilon > 0$ be given.
Find a finite $K = K_\epsilon$,  points $x_1,\ldots, x_K$ in $[1/4,3/4]^d$ 
and $r = r_\epsilon > 0$ such that
(1) holds for any vector $(x'_1,\ldots,x'_{K})$ with
$x'_i \in  B(x_i, r)$, $1 \le i \le K$.  In our application, this is a locally stable Delaunay configuration with large dilation. 
Define $S_i = B(x_i, r)$. By choice of $r$, we can insure that
the $S_i$ are non-overlapping and completely contained in $[0,1]^d$.

Fix $a$ and $b$ as in Lemma~\ref{lem:2}.
Armed with the sets $S_i$, and given any integer $N > 0$,
Lemma~\ref{lem:2} tells us that we can find
$\delta_0 > 0$,
$y_1,\ldots,y_N \in \RR^d$,
and constant $c > 0$
such that
$b \ge f(x_i) \ge a$, 
the $L_\infty$ distance between each $x_i$ and $x_j$
is more than $c$, and 
	$$
	{1 \over \delta^d \lambda (S_j)} 
	\int_{x_i + \delta S_j} f \in [ f(x_i)/2 , 2f(x_i) ] ,
        $$
for all $1 \le i \le N$, $0 \le j \le K$, $0 \le \delta \le \delta_0$,
	where $S_0 = [0,1]^d$.

Choose $\delta = 1/n^{1/d}$, and let $n$ be large enough to insure that $\delta \le \min(c, \delta_0)$.
For a set $A$, we define its cardinality by the counting measure,
$$
|A| = \sum_{i=1}^n \IND{X_i \in A},
$$
where $X_1,\ldots,X_n$ are the i.i.d.\ data drawn from $f$.
For $1 \le i \le N$, define the events
$$
E_i = \cap_{j=1}^K \left[ | x_i + \delta S_j | = 1 \right] \, \cap \left[ | x_i + \delta ( [0,1]^d - \cup_{1 \le j \le K} S_j )  | = 0 \right].
$$
It is clear that if one of the events $E_i$ happens, then
$$
L_n (X_1,\ldots,X_n) \ge L^* - \epsilon.
$$
Set $p_i = \PROB \{ E_i \}$ and $p_{ij} = \PROB \{ E_i \cap E_j \}$.
We have, by the second moment method~\cite{ce-abcl-52},
$$
\PROB \left\{ \cup_i E_i \right\}
\ge
{ \left( \sum_i  p_i \right)^2 
  \over
  \sum_i p_i + \sum_{i \not= j} p_{ij} } .
$$
It helps to consider the ratios
$$
{p_{ij} \over p_i p_j}.
$$
Define
$$
q_i (\ell) = \int_{x_i + \delta S_\ell} f , 1 \le \ell \le K,
$$
and
$q_i (0) = \int_{x_i + \delta [0,1]^d } f$.
Then by the multinomial formula,
$$
p_i = {n \choose 1,1,1,\cdots,1,0,n-K} q_i (1) \cdots q_i (K) \left( 1 - q_i (0) \right)^{n-K}
= {n! \over (n-K)!} \, q_i (1) \cdots q_i (K) \left( 1 - q_i (0) \right)^{n-K}
$$
and
$$
p_{ij} = {n! \over (n-2K)!} \, q_i (1) \cdots q_i (K) q_j (1) \cdots q_j (K) \left( 1 - q_i (0) - q_j (0) \right)^{n-2K}.
$$
Thus,
\begin{align*}
{p_{ij} \over p_i p_j}
&= {(n-K)! (n-K)! \over n! (n-2K)!} { \left( 1 - q_i (0) - q_j (0) \right)^{n-2K}
   \over \left( 1 - q_i (0) \right)^{n-K} \left( 1 - q_j (0) \right)^{n-K} } \\
&\le {(n-K)! (n-K)! \over n! (n-2K)!}  \left( 1 - q_i (0) - q_j (0) \right)^{-K} \\
&\le \left( 1 - 4b \delta^d \right)^{-K} \\
&= \left( 1 - 4b/n \right)^{-K} \\
&\le { 1 \over 1 - {4bK \over n} } \\
&\isdef \gamma 
\end{align*}
if $n > 4bK$.
Thus, we have
\begin{align*}
\PROB \left\{ \cup_i E_i \right\}
&\ge { \left( \sum_i  p_i \right)^2 \over \sum_i p_i + \gamma \sum_{i \not= j} p_i p_j }  \\
&\ge { \left( \sum_i  p_i \right)^2 \over \sum_i p_i + \gamma \left( \sum_i p_i \right)^2  }  \\
&= { 1 \over \gamma + \left( \sum_i  p_i \right)^{-1} } . 
\end{align*}
We have $\gamma \to 1$ as $n \to \infty$.
Furthermore, $\sum_i  p_i$ can be made as large as desired.
To see this,
note that $q_i (0) \le 2 f(x_i) \delta^d \le 2b/n $, $q_i (\ell) \ge (1/2) f(x_i) \delta^d \lambda (S_\ell) \ge (1/2) a \lambda (S_\ell)/n \isdef \xi/n $ for
$\ell > 0$ as each $S_\ell$ has the same positive volume.
Thus,
$$
\sum_i p_i \ge N {n! \over (n-K)!} (\xi/n)^K (1- 2b/n)^{n-K} 
\sim {N \xi^K e^{-2b} \over K!},
$$
which is as large as desired by choice of $N$.
Therefore,
$$
\lim_{n \to \infty} \PROB \{ L_n (X_1,\ldots,X_n) \ge L^* - \epsilon \} = 1. ~\square
$$
\medskip

\section{Open problems}
There remains a large gap between the upper and lower bounds on the dilation of the Delaunay triangulation, both for points in convex and general position.  We still believe that the true bound is near 1.6 in both cases; even in the convex case it would be interesting to reduce the upper bound.

\singlespace
\section* {Acknowledgments}
The authors are grateful to all three reviewers for their detailed reading and constructive comments. Over the course of several years, the first author thanks the following people for fruitful discussions on this topic: Sebastien Collette, Paz Carmi, Ferran Hurtado, John Iacono, Mark Keil, Stefan Langerman, Pat Morin, Jason Morrison, Michiel Smid, and Jorge Urrutia (who owes him coffee). The third author would like to thank Herman Haverkort and Elena Mumford for providing a stimulating working environment while working on this problem.

  \bibliographystyle {abbrv}
  \bibliography {refs}

\end {document}